\documentclass[prd,nofootinbib,preprint,showpacs,showkeys,superscriptaddress]{revtex4}
\usepackage[english]{babel}
\usepackage{amsmath,amssymb}
\usepackage[dvips]{graphicx}
\usepackage{ifthen,array}
\usepackage[sort&compress]{natbib}
\usepackage{amsfonts}
\usepackage{bbm}
\usepackage{axodraw}
\usepackage{enumerate}

\def\e6{$E(6)$}
\def\10{$SO(10)$}
\def\21{SU(2) $\otimes$ U(1) }
\newcommand{\lr}{$SU(2)_L \otimes SU(2)_R \otimes U(1)$}
\def\422{$SU(4) \otimes SU(2) \otimes SU(2)$}
\def\321{SU(3) $\otimes$ SU(2) $\otimes$ U(1)}

\def\vev#1{\left\langle #1\right\rangle}

\def\mathbf#1{\hbox{\bf #1}}
\def\textrm#1{\hbox{#1}}

\def\lsim{\raise0.3ex\hbox{$\;<$\kern-0.75em\raise-1.1ex\hbox{$\sim\;$}}}
\def\gsim{\raise0.3ex\hbox{$\;>$\kern-0.75em\raise-1.1ex\hbox{$\sim\;$}}}
\newcommand {\ignore}[1]{}

\newcommand{\AddrAHEP}{%
  AHEP Group, Instituto de F\'{\i}sica Corpuscular --
  C.S.I.C./Universitat de Val{\`e}ncia \\
  Edificio Institutos de Paterna, Apt 22085, E--46071 Valencia, Spain}
\newcommand{\AddrWU}{%
  Institut f\"ur Theoretische Physik, Universit\"at
  W\"urzburg, Am Hubland, D-97074~W\"urzburg, Germany}

\begin{document}
\baselineskip=1.1\baselineskip
\preprint{IFIC/04-25}

\vspace*{2cm} \title{ Enhanced lepton flavour violation \\ in the
  supersymmetric inverse seesaw model}

\author{F. Deppisch} \email{deppisch@physik.uni-wuerzburg.de}\affiliation{\AddrAHEP}\affiliation{\AddrWU}
\author{J.~W.~F.~Valle} \email{valle@ific.uv.es}\affiliation{\AddrAHEP}

\vspace*{1.0cm}

\begin{abstract}
  
  We discuss a supersymmetric inverse seesaw model in which lepton
  flavour violating decays can be enhanced either by flavour violating
  slepton contributions or by the non-unitarity of the charged current
  mixing matrix.  As an example we calculate \(Br(\mu\to e\gamma)\)
  taking into account both heavy lepton exchange as well as
  supersymmetric diagrams in a minimal supergravity framework.  We
  find that the for the same parameters the rate can be enhanced with
  respect to seesaw model expectations, with or without supersymmetry.

\end{abstract}


\maketitle

\section{Introduction}

A heroic effort dating back to over four decades has finally led
to the discovery of neutrino oscillations, through a combination
of solar, atmospheric, reactor and accelerator neutrino
experiments~\cite{fukuda:1998mi,ahmad:2002jz,eguchi:2002dm,Maltoni:2004ei}.
This has now firmly established the incompleteness of the Standard
Model of electroweak interactions, suggesting that lepton flavour
violation (LFV) may also take place in the charged lepton sector.

It is well known that within the framework of the standard
non-supersymmetric seesaw models of neutrino
masses~\cite{gell-mann:1980vs,yanagida:1979,schechter:1980gr,mohapatra:1981yp}
rare lepton flavour violating decays like \(\mu\to e\gamma\) are
indeed extremely rare, far beyond the sensitivity of any
foreseeable experiment. In these models the effective light
neutrino mass $m_\nu$ is inversely proportional to the scale of
the lepton number violating \21 singlet Majorana mass $M_R$,
$m_\nu \propto M_R^{-1}$, while typical LFV rates vanish with
higher powers of $M_R$.

There are two popular mechanisms where lepton flavour violation
can be greatly enhanced.  First, by supersymmetrizing the model,
in which case flavour violation in the neutrino Yukawa couplings
is automatically transmitted to charged leptons through charged
slepton and sneutrino
loops~\cite{hall:1986dx,borzumati:1986qx,barbieri:1994pv}, giving
a sizeable enhancement to the LFV decay rates.

The other possibility is to consider
variants~\cite{mohapatra:1986bd} of the seesaw scheme
characterized by a small effective lepton number violating
Majorana mass term $\mu$ so that $m_\nu \propto \mu$. Its
smallness may be technically natural, since as $m_\nu \to 0$ a
larger symmetry is achieved~\cite{'tHooft:1979bh}, namely lepton
number is restored and neutrinos become massless. Due to the fact
that in this case $m_\nu \to 0$ as $\mu \to 0$ this scheme may be
called ``inverse seesaw''.  The resulting LFV rates do not depend
at all on the magnitude of the lepton number violating scale
$\mu$, which can lie below the weak scale. Moreover, LFV processes
may take place even in the limit where lepton number is strictly
conserved~\cite{bernabeu:1987gr}.  Similarly, CP violation can
arise in the limit where the light neutrinos are strictly
massless~\cite{branco:1989bn,rius:1990gk}.  The model thus serves
to elucidate that, from a basic point of view, neutrino masses do
not play a fundamental role in generating flavour violating
processes, which in this case are mediated by the exchange of \21
singlet leptons, due to the structure of the electroweak charged
and neutral currents~\cite{schechter:1980gr}.
However, in contrast to the standard seesaw, in inverse seesaw schemes
the \21 singlet leptons need not be super-heavy, leading to highly
enhanced LFV rates irrespective of the massiveness of neutrinos and
irrespective of the existence of supersymmetric
states~\cite{bernabeu:1987gr,branco:1989bn,rius:1990gk,gonzalez-garcia:1992be,valle:1991pk}.

This paper is organized as follows.  In
Sec.~\ref{sec:inverse-sees-mech} we introduce the main features of
the model, while in Sec.~\ref{sec:renorm-group-evol} we give the
renormalization group evolution both of the neutrino sector as
well as the slepton sector.  In Sec.~\ref{sec:lept-flav-viol} we
discuss the relative importance of the two types of contribution
to \(l_i^-\to l_j^-\gamma\) decays and in
Sec.~\ref{sec:numerical-results} we give our numerical results and
summarize the findings in Sec. \ref{sec:conclusions}.


\section{Inverse seesaw mechanism}
\label{sec:inverse-sees-mech}

The particle content of left-handed leptons in the model extends
minimally that of the Standard Model, by the sequential addition of a
pair of two-component \21 singlet fermions, as follows
\begin{equation}
\left(\begin{array}{c}
   \nu_i \\
   e_i \\
\end{array}\right), e_i^c, \nu_i^c, S_i,
\end{equation}
with \(i\) a generation index running over \(1,2,3\). In addition
to the right-handed neutrinos characteristic of the standard
seesaw model, the inverse seesaw scheme contains an equal number
of gauge singlet neutrinos \(S_i\).  In the original formulation
of the model, these were superstring inspired \e6 singlets, in
contrast to the right-handed neutrinos, members of the spinorial
representation. A similar construction at the \lr\ level was
considered in Ref.~\cite{Wyler:1983dd}.

The model is characterized by the following symmetric \(9\times9\)
mass matrix $\mathcal{M}$ in the \(\nu,\nu^c,S\) basis,
\begin{equation}
\label{eqn:invSeesaw}
\mathcal{M}=\left(
   \begin{array}{ccc}
      0   & m_D^T & 0   \\
      m_D & 0     & M^T \\
      0   & M     & \mu
   \end{array}\right),
\end{equation}
where \(m_D\) and \(M\) are arbitrary \(3\times3\) complex
matrices in flavour space, whereas \(\mu\) is complex symmetric.
The matrix \(\mathcal{M}\) can be diagonalized by a unitary mixing
matrix \(U_\nu\),
\begin{equation}
   U_\nu^T \mathcal{M} U_\nu = \mathrm{diag}(m_i,M_4,...,M_9),
\end{equation}
yielding 9 mass eigenstates \(n_a\), three of them corresponding
to the observed light neutrinos with masses \(m_i\), plus the
three pairs of two-component leptons \((\nu^c_i,S_i)\) combining
to form three heavy quasi-Dirac leptons~\cite{valle:1983yw}.

The mass eigenstates are then related to the light neutrino
flavour states \(\nu_i\) via the unitary matrix $U_\nu$
\begin{equation}
   \nu_i = \sum_{a=1}^{9}(U_\nu)_{ia} n_a.
\end{equation}
which has been studied in earlier
papers~\cite{bernabeu:1987gr,branco:1989bn,rius:1990gk}. Assuming
\(m_D,\mu \ll M\) the diagonalization results in an effective Majorana
mass matrix for the light neutrinos~\cite{gonzalez-garcia:1989rw},
\begin{equation}
\label{eqn:lightNu}
    m_\nu = {m_D^T M^{T}}^{-1} \mu M^{-1} m_D,
\end{equation}
which may be estimated as
\begin{equation}
    \left(\frac{m_\nu}{0.1\mathrm{eV}}\right) =
        \left(\frac{m_D}{100\mathrm{GeV}}\right)^2
        \left(\frac{\mu}{1\mathrm{keV}}\right)
        \left(\frac{M}{10^4\mathrm{GeV}}\right)^{-2}
\label{eqn:lightNuNumeric},
\end{equation}
which vanish in the limit $\mu \to 0$ where lepton number
conservation is restored. In models where lepton number is
spontaneously broken by a vacuum expectation value
$\vev{\sigma}$~\cite{gonzalez-garcia:1989rw} one has $\mu =
\lambda \vev{\sigma}$.  For typical Yukawas $\lambda \sim 10^{-3}$
one sees that $\mu = 1$ keV corresponds to a low scale of L
violation, $\vev{\sigma} \sim 1$ MeV. Although such a low scale is
not protected by supersymmetry, this is not needed, as it is
already protected by gauge symmetry. Indeed, being a gauge
singlet, the smallness of $\vev{\sigma}$ is not destabilized by
gauge loops and is technically natural~\footnote{For very low
values this might lead to interesting signatures in neutrinoless
double beta decays~\cite{berezhiani:1992cd}.}.

In such an ``inverse seesaw'' scheme the three pairs of singlet
neutrinos have masses of the order of \(M\) and their admixture in
the light neutrinos is suppressed as \(\frac{m_D}{M}\).

In contrast, in the standard seesaw mechanism where the gauge singlet
neutrinos \(S_i\) are absent one would have
\begin{equation}
\label{eqn:Seesaw} \left(
   \begin{array}{cc}
      0   & m_D^T   \\
      m_D & M_R
   \end{array}\right),\qquad m_D \ll M_R \Rightarrow m_\nu = m_D^T M_R^{-1} m_D\,.
\end{equation}

Note that although \(M\) and \(M_R\) both take the role of a large
mass scale suppressing the light neutrino masses, their physical
meaning is quite different, the former being a Dirac mass (\(\Delta L
= 0\)) and the latter a Majorana mass (\(\Delta L = 2\)).  In contrast
to the mass \(M_R\) of the right-handed neutrinos of the standard
seesaw model, the mass \(M\) of our heavy leptons can be much smaller,
since the suppression in Eq.~(\ref{eqn:lightNuNumeric}) is quadratic
and since we have the small independent parameter \(\mu\)
characterizing the violation of lepton number. As a result the value
of \(M\) may be as low as the weak scale~\footnote{If light enough,
  these neutral leptons would be singly-produced in \(Z^0\)
  decays~\cite{Dittmar:1990yg,Gonzalez-Garcia:1990fb}, a possibility
  now ruled out by LEP~\cite{Abreu:1997pa}.}.

Without loss of generality one can assume \(\mu\) to be diagonal,
\begin{equation}
    \mu = \textrm{diag}\;\mu_i,
\end{equation}
and using the diagonalizing matrix \(U\) of the effective light
neutrino mass matrix \(m_\nu\),
\begin{equation}
    U^T m_\nu U = \textrm{diag}\;m_i,
\end{equation}
equation (\ref{eqn:lightNu}) can be written as
\begin{equation}
    \mathbf{1} = \textrm{diag}\;\sqrt{m_i^{-1}} \cdot U^T m_D^T {M^T}^{-1}
                 \cdot \textrm{diag}\;\sqrt{\mu_i} \cdot
                 \textrm{diag}\;\sqrt{\mu_i} \cdot M^{-1} m_D U
                 \cdot \textrm{diag}\;\sqrt{m_i^{-1}}.
\end{equation}
In the basis where the charged lepton Yukawa couplings are diagonal
the lepton mixing matrix is simply the rectangular matrix formed by
the first three rows of \(U_\nu\)~\cite{schechter:1980gr}.

In analogy to the standard seesaw mechanism \cite{Casas:2001sr} it is
thus possible to define a complex orthogonal matrix
\begin{equation}
\label{eqn:R}
    R = \textrm{diag}\;\sqrt{\mu_i} \cdot M^{-1} m_D U \cdot \textrm{diag}\;\sqrt{m_i^{-1}}
\end{equation}
with 6 real parameters.  Using \(R\), the neutrino Yukawa coupling
matrix \(Y_\nu = \frac{1}{v\sin\beta} m_D\) can be expressed as
\begin{equation}
\label{eqn:R}
    Y_\nu = \frac{1}{v\sin\beta} M \cdot \textrm{diag}\;\sqrt{\mu_i^{-1}} \cdot R
            \cdot \textrm{diag}\;\sqrt{m_i} \cdot U^\dagger,
\end{equation}
To simplify our discussion we make the assumption that the eigenvalues
of both \(M\) and \(\mu\) are degenerate and that \(R\) is real. This
allows us to easily compare our results with those obtained previously
in Ref.~\cite{Deppisch:2002vz,Deppisch:2003wt} for the case of the
standard seesaw mechanism.  The combination \(Y_\nu^\dagger L Y_\nu\)
which is responsible for flavour non-diagonal slepton mass terms in
Eqs.~(\ref{eq:rnrges1},\ref{eq:rnrges3}) is given by
\begin{equation}
\label{eqn:YLY}
    (Y_\nu^\dagger L Y_\nu)_{ij} =
        \frac{1}{v^2 \sin^2\beta}\frac{M^2}{\mu} \ln\frac{M_{GUT}}{M}
        (U \cdot \textrm{diag}\;m_i \cdot U^\dagger)_{ij}.
\end{equation}
This should be compared with the elements of the light neutrino mass
matrix
\begin{equation}
\label{eqn:m_ij}
    (m_\nu)_{ij} = (U \cdot \textrm{diag}\;m_i \cdot U^T)_{ij}
\end{equation}
The physical consequence of the simplifications we use is that the
pattern of LFV transmitted to the sleptons is closely correlated
to that of the light neutrino sector. The only difference being
the roles of CP violating phases in (\ref{eqn:YLY}) and
(\ref{eqn:m_ij}) that may be present in \(U\). In the case of CP
conservation which we will consider, the correlation is exact.

In the standard supersymmetric seesaw mechanism the flavour
non-diagonal slepton mass terms would be (in an analogous
approximation, i.e. \(M_i=M_R\) and real \(R\) matrix)
\begin{equation}
\label{eqn:YLY-standard-seesaw}
    (Y_\nu^\dagger L Y_\nu)_{ij} =
        \frac{1}{v^2 \sin^2\beta} M_R \ln\frac{M_{GUT}}{M_R}
        (U \cdot \textrm{diag}\;m_i \cdot U^\dagger)_{ij},
\end{equation}
as compared to Eq.~\ref{eqn:YLY}.


\section{Renormalization group evolution}
\label{sec:renorm-group-evol}

\subsection{ The neutrino sector}
\label{sec:neutrino-sector}

In what follows it will be sufficient for us to confine ourselves to
the simpler case where heavy neutrino masses are degenerate.  Below
the scale of \(M\), the one-loop renormalization group equation (RGE)
for the effective neutrino mass matrix in the MSSM is given by
\cite{Ellis:1999my}
\begin{equation}
\label{eqn:diffMnu}
    \frac{d}{dt}m_\nu =
        \frac{1}{16 \pi^2} \left(\left(-6g_2^2-\frac{6}{5}g_1^2
        + Tr(6 Y_U^\dag Y_U)\right)
        m_\nu + \left((Y_l^\dag Y_l)m_\nu
        + m_\nu (Y_l^\dag Y_l)^T\right)\right),
\end{equation}
with the U(1) and SU(2) gauge couplings \(g_1\) and \(g_2\), and the
Yukawa coupling matrices \(Y_U\) and \(Y_l\) for the charge
\(\frac{2}{3}\)-quarks and charged leptons, respectively~\footnote{The
  corresponding evolution equations for \(g_{1,2}\), \(Y_{U}\) and
  \(Y_{l}\) can be found in \cite{deBoer:1994dg}.}. This RGE is linear
in \(m_\nu\) and can thus be solved analytically \cite{Ellis:1999my}
as
\begin{equation}
    m_\nu(t) = I(t) m_\nu(0) I(t), \quad t=\ln \left(\frac{\mu}{M_Z}\right)\,.
\end{equation}
Since the evolution is dominated by the gauge and third generation
Yukawa couplings one obtains, to a good approximation:
\begin{equation}
    I(t)=I_g I_t \cdot \textrm{diag}\left(1,1,I_\tau\right)
\end{equation}
with
\begin{eqnarray}
    I_g(t)     &=& \exp \left( \frac{1}{16\pi^2}\int_0^t (-3g_2^2-\frac{3}{5}g_1^2)dt'\right), \\
    I_t(t)     &=& \exp \left( \frac{1}{16\pi^2}\int_0^t 3|Y_t|^2 dt'\right), \\
    I_\tau(t)  &=& \exp \left( \frac{1}{16\pi^2}\int_0^t |Y_\tau|^2dt'\right).
\end{eqnarray}
Above the scale \(M\), which can be substantially lower than the
corresponding heavy lepton scale in standard seesaw schemes, the
evolution of the neutrino Yukawa matrix \(Y_\nu\) is governed
by~\cite{Casas:2001sr}
\begin{equation}
\label{eqn:diffYnu}
    \frac{d}{dt}Y_\nu = \frac{1}{16 \pi^2}Y_\nu\left(\left(-3g_2^2-\frac{3}{5}g_1^2 + Tr(3 Y_U^\dag Y_U
                        + Y_\nu^\dag Y_\nu)\right)\mathbf{1} + Y_l^\dag Y_l + 3Y_\nu^\dag Y_\nu\right).
\end{equation}


\subsection{The slepton sector}
\label{sec:slepton-sector}

Having evolved the neutrino Yukawa couplings to the unification scale
\(M_{GUT}\), one can now run the slepton mass matrix from \(M_{GUT}\)
to the electroweak scale assuming the mSUGRA universality conditions
on the soft SUSY breaking terms \(m^2_L\) (left-handed slepton
doublets), \(m^2_R\) (charged right-handed slepton singlets),
\(m^2_{\tilde N}\) (right-handed sneutrino singlets), \(A_e\)
(trilinear couplings analogous to \(Y_e\)) and \(A_\nu\) (trilinear
couplings analogous to \(Y_\nu\)) at \(M_{GUT}\):
\begin{equation}
    m^2_L =  m^2_R = m^2_{\tilde N} = m_{0}^{2}\mathbf{1},
    \quad A_e = A_{0}Y_e, \quad A_\nu=A_0 Y_\nu \label{mSUGRAcond},
\end{equation}
where \(m_{0}\) is the common scalar mass and \(A_{0}\) the common
trilinear coupling.  For definiteness we adopt in our following
analysis, the mSUGRA benchmark scenario SPS1a proposed in
\cite{allanach:2002nj}, described by
\begin{equation}
    m_0=100\textrm{ GeV}, m_{1/2}=250\textrm{ GeV},
    A_0=-100\textrm{ GeV}, \tan\beta=10, \mu > 0,
\end{equation}
where \(\mu\) is here the SUSY Higgs-mixing parameter. In general, the
charged slepton \((\textrm{mass})^2\) matrix has the form:
\begin{equation}
\label{ch_slepton_mass_mat}
    m_{\tilde l}^2=\left(
                   \begin{array}{cc}
                         m^2_{\tilde{l}_{L}}    & (m^{2}_{\tilde{l}_{LR}})^\dagger \\
                         m^{2}_{\tilde{l}_{LR}} & m^{2}_{\tilde{l}_{R}}
                   \end{array}
                   \right),
\end{equation}
where \(m^2_{\tilde{l}_{L}}\), \(m^{2}_{\tilde{l}_{R}}\) and
\(m^{2}_{\tilde{l}_{LR}}\) are \(3\times3\) matrices,
\(m^2_{\tilde{l}_{L}}\) and \(m^{2}_{\tilde{l}_{R}}\) being hermitian.
The matrix elements are given by
\begin{eqnarray}
    (m^2_{\tilde{l}_L})_{ab}     &=& (m_{L}^2)_{ab} + \delta_{ab}\left(m_{l_a}^2 +
                                     m_Z^2 \cos(2\beta) \left(-\frac{1}{2}
                                                              +\sin^2\theta_W \right)\right)
                                     \label{mlcharged} \\
    (m^2_{\tilde{l}_{R}})_{ab}   &=& (m_{R}^2)_{ab} + \delta_{ab}(m_{l_a}^2 -
                                     m_Z^2 \cos(2\beta)\sin^2\theta_W) \label{mrcharged} \\
    (m^{2}_{\tilde{l}_{LR}})_{ab}&=& (A_\nu)_{ab} v \cos\beta-\delta_{ab}m_{l_a}\mu\tan\beta.
\end{eqnarray}
Applying the mSUGRA conditions (\ref{mSUGRAcond}) at \(M_{GUT}\) and
performing the evolution to \(M_{Z}\) the SUSY breaking terms can be
expressed as:
\begin{eqnarray}
    m_{L}^2 &=& m_0^2\mathbf{1} + (\delta m_{L}^2)_{\textrm{\tiny MSSM}} +
                \delta m_{L}^2 \label{left_handed_SSB} \\
    m_{R}^2 &=& m_0^2\mathbf{1} + (\delta m_{R}^2)_{\textrm{\tiny MSSM}} +
                \delta m_{R}^2 \label{right_handed_SSB}\\
    A_e     &=& A_0 Y_e+\delta A_{\textrm{\tiny MSSM}}+\delta A \label{A_SSB},
\end{eqnarray}
with \((\delta m^{2}_{L,R})_{\textrm{\tiny MSSM}}\) and \((\delta
A)_{\textrm{\tiny MSSM}}\) denoting the usual flavour diagonal
MSSM renormalization group corrections \cite{deBoer:1994dg}.  In
addition, the presence of right-handed neutrinos radiatively
induces flavour off-diagonal terms denoted by \(\delta m_{L,R}^2\)
and \(\delta A\). In the leading-log approximation these terms are
given as~\cite{Hisano:1999fj}
\begin{eqnarray}
    \delta m_{L}^2 &=&
        -\frac{1}{8 \pi^2}(3m_0^2+A_0^2)(Y_\nu^\dag L Y_\nu)\label{eq:rnrges1}\\
    \delta m_{R}^2 &=& 0 \label{delta_m_R} \label{eq:rnrges2}\\
    \delta A       &=& -\frac{3 A_0}{16\pi^2}Y_e \cdot (Y_\nu^\dag L Y_\nu)
\label{eq:rnrges3}
\end{eqnarray}
with
\begin{equation}
    L = \textrm{diag}\left(\ln\frac{M_{GUT}}{M_i}\right).
\end{equation}
Finally, the physical charged slepton masses are then found by
diagonalizing (\ref{ch_slepton_mass_mat}) using the \(6\times6\)
unitary matrix \(U_{\tilde l}\):
\begin{equation}
    U_{\tilde l}^{\dagger} m_{\tilde l}^2 U_{\tilde l} =
        \textrm{diag}(m_{\tilde l_1}^2,...,m_{\tilde l_6}^2).
\end{equation}
Correspondingly, the slepton mass eigenstates are expressed in terms
of the gauge eigenstates by
\begin{equation}
    \tilde l_a = (U_{\tilde l}^{*})_{ia}\tilde l_{Li}
                 + (U_{\tilde l}^{*})_{(i+3)a} \tilde l_{Ri},
                 \quad\quad a=1,...,6; \; i=e,\mu,\tau.
\end{equation}
Similarly to (\ref{mlcharged}), the \(6\times6\) (mass\()^2\)
matrix of the SUSY partners of the left- and right-handed
neutrinos is given by
\begin{equation}
\label{ch_sneutrino_mass_mat}
    m_{\tilde l}^2 =
    \left(\begin{array}{cc}
        m^2_{\tilde{\nu}_{L}}    & (m^{2}_{\tilde{\nu}_{LR}})^\dagger \\
        m^{2}_{\tilde{\nu}_{LR}} & m^{2}_{\tilde{\nu}_{R}}
    \end{array}\right),
\end{equation}
%
%
with
\begin{eqnarray}
    (m^2_{\tilde{\nu}_L})_{ab}      &=&
        (m_{L}^2)_{ab}+\frac{1}{2}\delta_{ab}m_Z^2\cos(2\beta) + (m_\nu)_{ab}^2 \\
    (m^2_{\tilde{\nu}_{R}})_{ab}    &=&
        (m_{\tilde N}^2)_{ab} + M_{ab}^2 \\
    (m^{2}_{\tilde{\nu}_{LR}})_{ab} &=&
        v\cos\beta(A_\nu)_{ab} - \mu\cot\beta(m_D)_{ab}.
\end{eqnarray}

The sneutrino mass matrix is diagonalized by a unitary \(6\times
6\) matrix \(U_{\tilde\nu}\),
\begin{equation}
    U_{\tilde\nu}^\dagger m_{\tilde\nu}^2 U_{\tilde\nu} = \textrm{diag}(m_{\tilde\nu_1}^2,
                                                          ...,m_{\tilde\nu_6}^2),
\end{equation}
which, like \(U_\nu\), can contain sizeable mixings between \21
isodoublet and isosinglet sneutrinos.  In contrast to the standard
seesaw where the right-handed neutrinos (and thus sneutrinos) are
extremely heavy and can be safely neglected in low energy processes,
this is not the case in our ``inverse seesaw'' model.  Indeed, the
presence of the gauge singlet superfields \(S_i\) is crucial in this
model in causing a big enhancement in the LFV rates. Note however that
the scalars present in Eq.~(\ref{ch_sneutrino_mass_mat}) can be
neglected in calculating the LFV decay rates.


\section{Lepton flavour violating decays: \(l_i^-\to l_j^-\gamma\)}
\label{sec:lept-flav-viol}

The effective Lagrangian for \(l_i^-\to l_j^-\gamma\) may be written
generically as
\begin{equation}
\mathcal{L}_{eff}=\frac{e}{2}\bar{l}_{j}\sigma_{\alpha \beta}F^{\alpha \beta}\left(A_{L}^{ij}P_{L}+
A^{ij}_{R}P_{R}\right)l_{i},\label{Leff}
\end{equation}
where \(F^{\alpha\beta}\) is the electromagnetic field strength
tensor,
\(\sigma_{\alpha\beta}=\frac{i}{2}\left[\gamma_{\alpha},\gamma_{\beta}\right]\)
and \(P_{R,L}=\frac{1}{2}(1\pm \gamma_{5})\) are the helicity
projection operators.  The coefficients \(A^{ij}_{L,R}\) are
determined by the relevant gauge theory Feynman diagrams.
The decay rate for \(l_{i}^{-} \to l_{j}^{-}\gamma\) that follows from
(\ref{Leff}) can be expressed as~\cite{Hisano:1996cp}
\begin{equation}
    \Gamma\left(l_i^-\to l_j^-\gamma\right) = \frac{\alpha}{4}m^{3}_{l_i}
    \left(\left| A_L^{ij}\right|^2 + \left|A_R^{ij}\right|^2\right)\label{dec_ij}.
\end{equation}

\subsection{Heavy lepton contribution}
\label{sec:heavy-lept-contr}

The contribution to the decay \(l_i^-\to l_j^-\gamma\) arising from
the admixture of the heavy neutrinos in the left-handed charged
current \21 weak interaction~\cite{schechter:1980gr} exists both for
the seesaw scheme as well as for the inverse seesaw model. One finds
that the branching ratio is given as \cite{Ilakovac:1994kj}
\begin{equation}
    \Gamma(l_i^-\to l_j^-\gamma) =
        \frac{\alpha_W^3 s_W^2}{256\pi^2}
        \left(\frac{m_{l_i}}{M_W}\right)^4 \frac{m_{l_i}}{\Gamma_i} |G_{ij}|^2,
\end{equation}
where
\begin{equation}
    G_{ij} = \sum_{k}(U_\nu)_{ik}^* (U_\nu)_{jk} \,
        G_\gamma\!\left(\frac{M_{N_k}^2}{M_W^2}\right),
    \quad G_\gamma(x) =
    -\frac{2x^3+5x^2-x}{4(1-x)^2}-\frac{3x^3}{2(1-x)^4}\ln x,
\end{equation}
\(\Gamma_i\) is the total decay rate of lepton \(i\) and \(U_\nu\)
is the matrix describing the diagonalization of the neutrino mass
matrix in the seesaw scheme under consideration.  As noted in
\cite{bernabeu:1987gr} within the framework of the inverse \21
seesaw mechanism the decay \(l_i^-\to l_j^-\gamma\) occurs in the
limit of lepton number conservation where the light neutrinos
become massless. The rate is correspondingly enhanced with respect
to that of the simplest seesaw scheme, as it is not suppressed by
the smallness of neutrino masses.
%

\subsection{Supersymmetric contribution}
\label{sec:supersymm-contr}

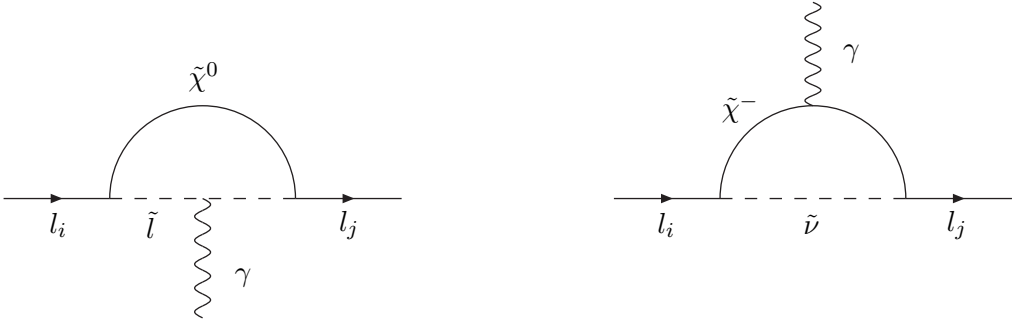
\begin{figure}
\centering
\begin{picture}(400,135)(-200,-50)
    \ArrowLine(-190,0)(-150,0)
    \Text(-170,-10)[]{\(l_{i}\)}
    \DashLine(-150,0)(-80,0){5}
    \Text(-135,-10)[]{\(\tilde{l}\)}
    \Photon(-115,0)(-115,-45){3}{5}
    \Text(-100,-30)[]{\(\gamma\)}
    \ArrowLine(-80,0)(-40,0)
    \Text(-60,-10)[]{\(l_{j}\)}
    \CArc(-115,0)(35,0,180)
    \Text(-115,46)[]{\(\tilde{\chi}^{0}\)}
    \ArrowLine(40,0)(80,0)
    \DashLine(80,0)(150,0){5}
    \ArrowLine(150,0)(190,0)
    \Text(60,-10)[]{\(l_{i}\)}
    \Text(115,-10)[]{\(\tilde{\nu}\)}
    \Photon(115,35)(115,74){3}{5}
    \Text(130,55)[]{\(\gamma\)}
    \Text(170,-10)[]{\(l_{j}\)}
    \CArc(115,0)(35,-360,-180)
    \Text(88,34)[]{\(\tilde{\chi}^{-}\)}
\end{picture}
\caption{Supersymmetric diagrams for \(l_{i}^{-} \to
l_{j}^{-}\gamma\)}\label{lfv_lowenergydiagrams}
\end{figure}

The coefficients \(A^{ij}_{L,R}\) of the effective Lagrangian for
\(l_i^-\to l_j^-\gamma\) in Eq.~(\ref{Leff}) in the MSSM have been
given in Ref.~\cite{Carvalho:2001ex}. They are determined by the
photon penguin diagrams shown in Fig.~\ref{lfv_lowenergydiagrams} with
charginos/sneutrinos or neutralinos/charged sleptons circulating in
the loop.
The superscript \(c\,(n)\) refers to the chargino (neutralino)
diagram of Fig.~\ref{lfv_lowenergydiagrams}, while the flavour
indices are omitted.  Because \(m_{l_{i}} \gg m_{l_{j}}\) and
\(m^{2}_{\tilde{l}_{R}}\) is diagonal (see
(\ref{mrcharged},\ref{right_handed_SSB},\ref{delta_m_R})), one has
\(A_{R}\gg A_{L}\) \cite{Casas:2001sr,Hisano:1999fj}.  The
dominant amplitudes in (\ref{dec_ij}) are approximately given by
\begin{eqnarray}
    A^c_R &\approx&
        \frac{1}{32\pi^2}\frac{g^2_2 m_{l_i}}{\sqrt{2}m_W\cos\beta}
        \sum_{a=1}^2\sum_{k=1}^6
            \frac{m_{\tilde\chi^-_a}}{m_{\tilde\nu_k}^2}(O_R)_{a1}(O_L)_{a2}
            (U_{\tilde\nu}^*)_{jk}(U_{\tilde\nu})_{ik}\nonumber\\
        & & \times\frac{-3+4r^c_{ak}-(r^c_{ak})^2-2\ln r^c_{ak}}{(1 - r^c_{ak})^3}
        \label{domchar} \\
    A^n_R &\approx&
        -\frac{1}{32\pi^2} g^2_2\tan\theta_W
        \sum_{a=1}^4\sum_{k=1}^6
            \frac{m_{\tilde\chi^0_a}}{m_{\tilde l_k}^2}
            (O_N)_{a1}((O_N)_{a2}+(O_N)_{a1}\tan\theta_W)\nonumber\\
        & & \times (U_{\tilde l}^*)_{jk}(U_{\tilde l})_{(i+3)k}
            \frac{1-(r^n_{ak})^2+2r^n_{ak}\ln r^n_{ak}}{(1-r^n_{ak})^3}
        \label{domneut}
\end{eqnarray}
with
\begin{equation}
    r^c_{ak} = \left(\frac{m_{\tilde\chi^-_a}}{m_{\tilde\nu_k}}\right)^2, \qquad
    r^n_{ak} = \left(\frac{m_{\tilde\chi^0_a}}{m_{\tilde l_k}}\right)^2,
\end{equation}
the chargino diagonalization matrices \(O_{L},O_{R}\) and the
neutralino diagonalization matrix \(O_{N}\). The mass eigenvalues of
the charginos and neutralinos are denoted by
\(m_{\tilde{\chi}^{-}_{a}}\) and \(m_{\tilde{\chi}^{0}_{a}}\),
respectively. It is worth noting that the sum in (\ref{domchar}) runs
over both left- and right-handed sneutrinos. The right-handed
sneutrinos will yield a sizeable contribution for small \(M\) whereas
the left-handed ones and the charged sleptons become significant for
larger \(M\).
%
%
The contribution due to flavour non-diagonal slepton mass terms is
roughly given by
\begin{equation}
\label{eqn:Br_approx}
   Br(l_i\to l_j\gamma) \approx \alpha^3 \tan^2\beta \left(\frac{m_{l_i}}{\tilde m}\right)^4 \frac{m_{l_i}}{\Gamma_i}
                                \left|\frac{(\delta m_L^2)_{ij}}{\tilde m^2}\right|^2
\end{equation}
where \(\tilde m\) is the approximate mass scale of SUSY particles in
the loops.

The numerical calculations discussed later are performed with the full
expressions for \(A_{L}^{c,n}\) and \(A_{R}^{c,n}\), which can be
found in \cite{Hisano:1996cp} and \cite{Okada:1999zk}.

\section{Numerical Results}
\label{sec:numerical-results}

\begin{figure}[t]
\centering
\includegraphics[clip,width=0.45\linewidth]{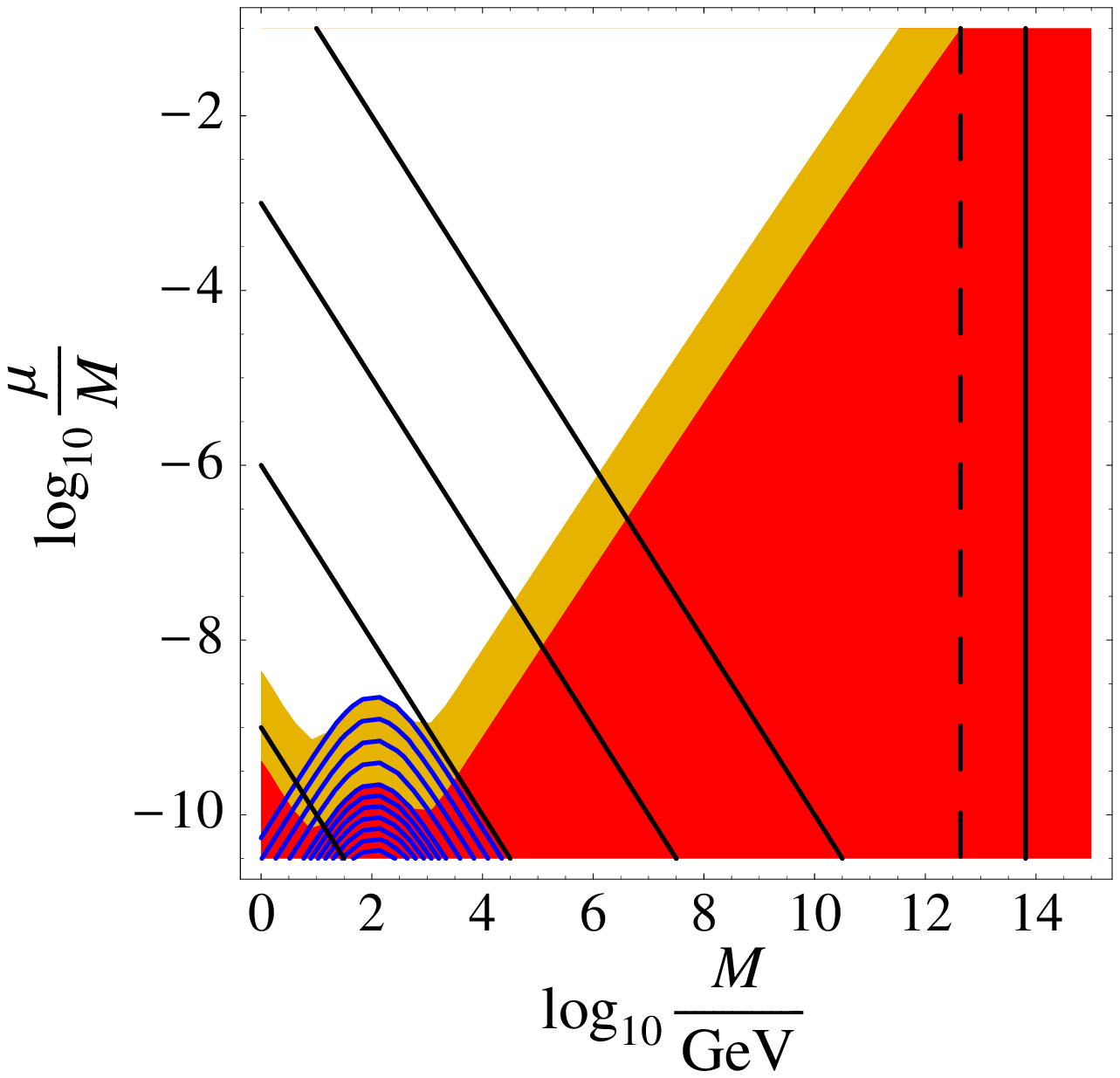}
\includegraphics[clip,width=0.45\linewidth]{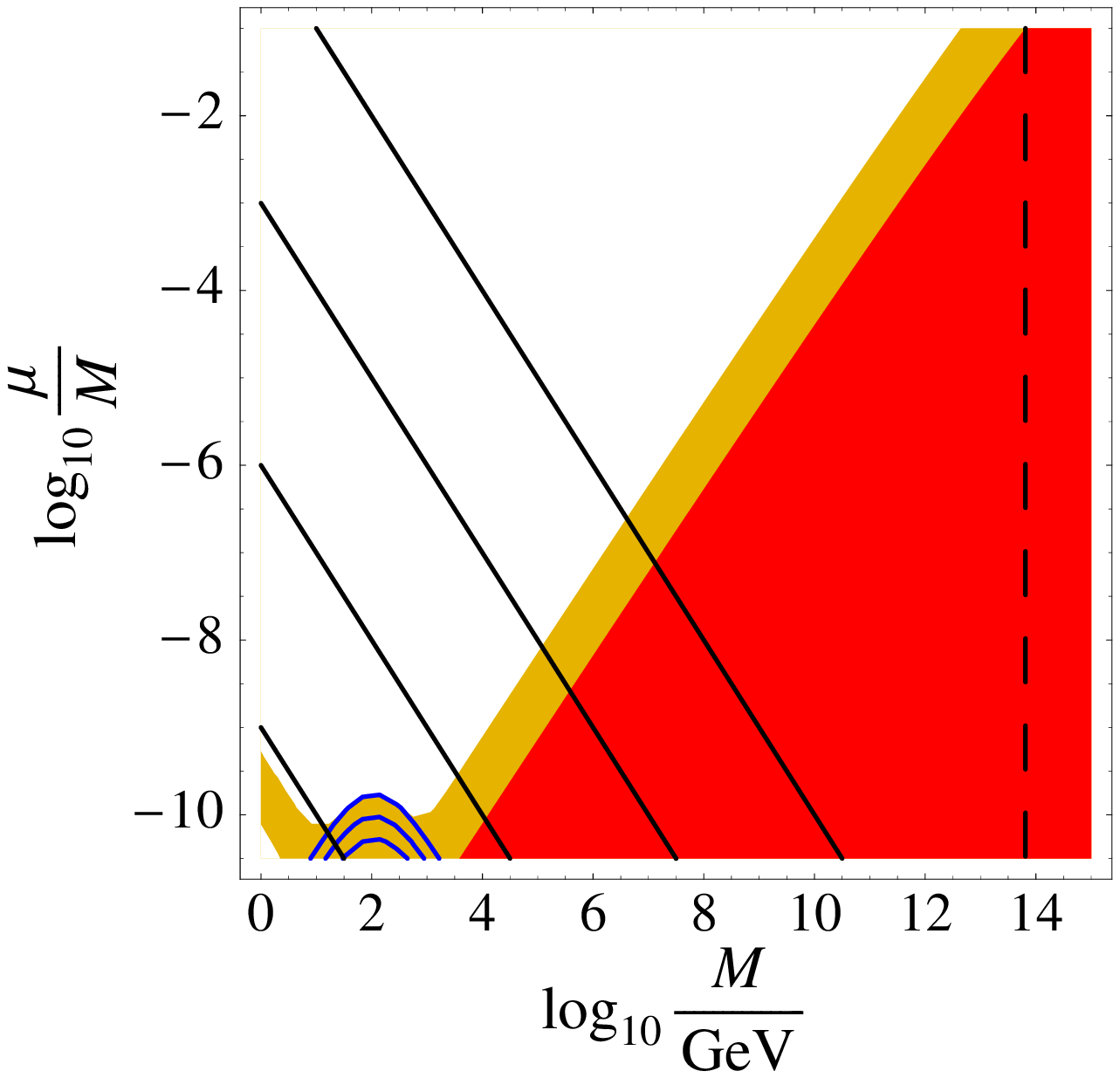}
     \caption{Contours of \(Br(\mu\to e\gamma)\) in the
       \((M,\frac{\mu}{M})\)-plane (logarithmic scales) for
       hierarchical light neutrinos with \(m_1=0\)~eV (left panel) and
       for degenerate light neutrinos with \(m_1=0.3\)~eV (right
       panel). The dark (light) area is excluded for \(Br(\mu \to
       e \gamma) < 10^{-11} (10^{-13})\). The blue contours in the
       lower left depict the contribution from neutral heavy leptons
       only. The diagonal lines show contours of constant \(\mu=
       1,10^{-3},10^{-6},10^{-9}\)~GeV (top to bottom).  The vertical
       lines are contours of \(Br(\mu\to e\gamma)\) in the standard
       SUSY seesaw (see text).}
     \label{fig:Br_M_mu}
\end{figure}

Low energy neutrino
experiments~\cite{fukuda:1998mi,ahmad:2002jz,eguchi:2002dm} now
provide substantial information on the light neutrino masses and
lepton mixing matrix.  This information has to be evolved to the
unification scale in order to calculate the slepton mass corrections.
Our numerical calculation is performed as follows.
We fix the light neutrino sector by using the latest global
fit~\cite{Maltoni:2004ei} for the neutrino oscillation parameters,
neglecting possible CP phases to which current data are insensitive.
The Yukawa coupling \(Y_\nu\) is then calculated via
Eq.~(\ref{eqn:R}). The result for \(Br(\mu\to e\gamma)\) in the case
of hierarchical neutrino masses, \(m_1 = 0\)~eV, is shown in
Fig.~\ref{fig:Br_M_mu} (left panel) as a contour plot in the
\((M,\frac{\mu}{M})\)-plane, with \(\mu < 0.1 M\) on the whole
plane~\footnote{ The diagonal lines depict contours of constant
  \(\mu\) in the inverse seesaw case, for easier reading.  }. The dark
(red) area is excluded by the current experimental limit \(Br(\mu\to
e\gamma) < 1.2 \cdot 10^{-11}\) while the light (brown) band shows the
additional sensitivity aimed at the PSI experiment \(Br(\mu\to
e\gamma) < 1.5 \cdot 10^{-13}\).
For large \(M\), the RGE-induced flavour violating slepton
contributions are clearly dominant due to the
\(\left(\frac{M}{\mu/M}\right)^2 \ln^2\frac{M_{GUT}}{M}\) dependence
from Eqs.~(\ref{eqn:YLY}) and (\ref{eqn:Br_approx}). It is only below
\(M \approx 10^3\)~GeV where the admixture of the heavy singlet
neutrino (and sneutrino) states becomes significant and dominant. In
order to guide the eye we have also included the corresponding
contours for the standard SUSY seesaw $\mu\to e\gamma$ decay branching
ratio at $10^{-11}$ and $10^{-13}$ (dashed), that would arise from
taking $M_R = M$, as indicated by the vertical lines on the right
side. These are excatly vertical, as the seesaw does not contain the
parameter $\mu$.  We see that for given $M_R = M$ value the rate can
be enhanced with respect to seesaw model expectations, with or without
supersymmetry. 

The curved (blue) contours peaked at slightly above M=100~GeV
correspond to the non-supersymmetric isosinglet neutral heavy lepton
contribution, considered
in~\cite{bernabeu:1987gr,gonzalez-garcia:1992be,Ilakovac:1994kj}.  The
difference is that now we take into account the neutrino masses
indicated by neutrino oscillation data. This actually has no impact,
as the violation of flavour in this case arises mainly from the
isosinglet neutral heavy lepton contribution, which in this model is
essentially unrelated to the light neutrino masses, due to the freedom
in choosing the value of $\mu$. On the other hand, in contrast to the
above works we explicitly correlate the LFV decay rate to the neutrino
oscillation mixing angles. Finally, the rise on the very far left of
the plots corresponds the the contribution in the supersymmetric loops
arising from \21 singlet scalar neutrinos.

The right panel in Fig.~\ref{fig:Br_M_mu} shows the analogous plot
for quasi-degenerate neutrino masses, \(m_1=0.3\)~eV, where one
can see that \(Br(\mu\to e\gamma)\) is suppressed by roughly one
order of magnitude as compared to the hierarchical neutrino case.
This is because the flavour non-diagonal elements of the light
neutrino mixing matrix (which are ultimately the source of LFV,
both RGE- and non-universality-induced) are suppressed as
\(\frac{\sqrt{\Delta
    m_{ij}^2}}{m_1}\) for large \(m_1\) 



\section{Conclusions}
\label{sec:conclusions}

We have discussed a supersymmetric inverse seesaw model in which
lepton flavour violating decays can be enhanced either by flavour
violating slepton contributions or due to heavy lepton exchange.
In contrast to the standard seesaw scheme, the \21 isosinglet
heavy leptons present in this model can be relatively light and
contribute significantly to LFV processes, irrespective of the
magnitude of neutrino masses and irrespective of the
supersymmetric contributions.
We have considered both types of contributions focusing on their
differences with respect to the standard supersymmetric seesaw scheme.
As an example we have calculated \(Br(\mu\to e\gamma)\) taking
into account both neutral heavy lepton as well as supersymmetric
diagrams in a minimal supergravity framework. Clearly, additional
LFV processes such as $\mu \to 3 e$, $\mu \to e$ conversion in
nuclei, as well as LFV $\tau$ decays can be considered
generalizing to the supersymmetric case the analysis presented in
\cite{gonzalez-garcia:1992be,Ilakovac:1994kj}.

\section*{Acknowledgements}

This work was supported by the Spanish grant BFM2002-00345, by the
European Commission Human Potential Programme RTN network
HPRN-CT-2000-00148 and by the European Science Foundation network
grant N.86.  FD was supported by the EU Research Training Site
contract HPMT-2000-00124.  We thank Martin Hirsch and Heinrich
P\"as for useful discussions.


\end{document}